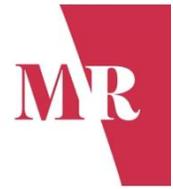

*Research Article*

# Disagreement as a way to study misinformation and its effects

*Experts consider misinformation a significant societal concern due to its associated problems like political polarization, erosion of trust, and public health challenges. However, these broad effects can occur independently of misinformation, illustrating a misalignment with the narrow focus of the prevailing misinformation concept. We propose using disagreement—conflicting attitudes and beliefs—as a more effective framework for studying these effects. This approach, for example, reveals the limitations of current misinformation interventions and offers a method to empirically test whether we are living in a post-truth era.*

Authors: Damian Hodel (1), Jevin D. West (1)
Affiliations: (1) Center for an Informed Public, Information School, University of Washington, USA

## Research questions

- What are the key problems attributed to misinformation (i.e., misinformation effects)?
- What are the limitations of the prevailing misinformation definitions and theories when studying these effects?
- Can disagreement serve as a more effective framework for analyzing, mitigating, and quantifying these misinformation effects?

## Essay summary

- We identify key limitations of the prevailing misinformation concept and propose disagreement as a more effective framework, drawing on literature from related disciplines.
- While misinformation concerns false and misleading information at the individual level, its effects are often shaped by normative, non-epistemic factors like identity and values, and can manifest at the societal level.
- Identifying misinformation is somewhat subjective, complicating automatic measurement and introducing a conflict of interest due to misinformation's negative connotation.
- In alignment with misinformation effects, disagreement is driven by normative factors and can occur at both individual and societal levels.

---

- Disagreement is necessary for misinformation effects, but misinformation is not.
- The disagreement framework reveals limitations of current intervention strategies.
- Disagreement can be identified without human judgment, enabling automated measurement of misinformation effects, as we demonstrate with two datasets comprising letters to the editor and Twitter posts.
- Measurement of disagreement in *The New York Times* letters to the editor from 1950 to 2022 reveals a rise of disagreement since 2006, indicating that misinformation effects may have increased in the last 20 years.
- Compared to the misinformation concept, disagreement is a more effective and efficient framework for studying misinformation effects because it offers better explanations for those effects, improves the development of targeted interventions, and provides a quantifiable method for evaluating the interventions.

## Implications

Experts and the public both consider misinformation a significant societal concern (Altay, Berriche, Heuer, et al., 2023; Ecker et al., 2024; McCorkindale, 2023), yet it remains a vague concept with inconsistent definitions (Adams et al., 2023; Altay, Berriche, & Acerbi, 2023; Southwell et al., 2022; Vraga & Bode, 2020) and a weak relationship with the societal or individual problems studied (Altay, Berriche, & Acerbi, 2023). These problems include polarization, erosion of institutions, problematic behavior, and individual and public health issues (Adams et al., 2023; Rocha et al., 2023; Tay et al., 2024). We refer to these problems as *misinformation effects* to indicate that they are assumed to be the effects of the spread of misinformation.[2] The gaps in prevailing definitions and theories of misinformation (Pasquetto et al., 2024) impact the analysis of misinformation effects (Altay, Berriche, & Acerbi, 2023), the development of intervention strategies (Aghajari et al., 2023), and the quantification of effects as a way to evaluate interventions (Southwell et al., 2022). We propose here disagreement as a more effective framework for studying these effects.

Given the variety of definitions for misinformation (Adams et al., 2023; Krause et al., 2022), we focus on the prevailing conceptualization: false and misleading information about factual matters that—whether intentionally or unintentionally—leads to or reinforces false beliefs (e.g., Altay, Berriche, Heuer, et al., 2023; Humprecht et al., 2020; Kennedy et al., 2022; Lazer et al., 2018; Southwell et al., 2022), serving as both a cause and effect of the aforementioned problems (see Figure 1).[3] While misinformation focuses on factual matters consumed at the individual level, the misinformation effects involve and are shaped by normative and societal factors beyond misinformation itself, such as opinions, values, epistemic positions, social norms, and ingroup and outgroup identity (see Figure 2).[4] In particular, the inclusion of misleading information makes identifying misinformation a highly context-sensitive task, often relying on a weak form of objectivity that can approach pure subjectivity (Krause et al., 2022; Uscinski, 2023). To clarify, we stand firmly behind scientific objectivity, viewing objectivity as epistemic rather than as metaphysical absolute value, borrowing from Frost-Arnold's (2023) definition: "a measurement of the degree to which bias is managed within a community of believers" (see also Longino, 2002). This subjectivity, combined with the conceptual misalignment, results in four key limitations when studying misinformation effects.

---

[2] Based on our analysis, we should refer to these as *disagreement effects*. However, for readability, we will use *misinformation effects* throughout the paper.

[3] Therefore, we exclude research that examines types of misinformation that link to problems without individuals holding false beliefs, such as inherently harmful information (e.g., toxic or discriminatory language).

[4] We are aware that recent studies sometimes use broader analytical frameworks, as we discuss in Conceptual Contribution 1. However, the primary focus in misinformation research remains on mitigating the spread of false and misleading information (Aghajari et al., 2023; e.g.,Kozyreva et al., 2024). Therefore, this is the standard against which we compare our proposed approach.

First, since misinformation is only one of many interdependent factors influencing beliefs and other misinformation effects (Altay, Berriche, & Acerbi, 2023; Tay et al., 2024), the current analytical framework provides a limited explanation for these effects: Someone can hold false beliefs without consuming misinformation (Altay, Berriche, & Acerbi, 2023), and the prevalence of misinformation does not necessarily indicate the prevalence of its effects (Southwell et al., 2022). Second, this weak relationship with misinformation limits the development of effective intervention strategies to counter misinformation effects and their associated harm (Aghajari et al., 2023), with experts skeptical about the efficacy of interventions to misinformation as a way to mitigate its effects (Acerbi et al., 2022; Altay, Berriche, Heuer, et al., 2023; Murphy et al., 2023). Third, the subjectivity in identifying misinformation, along with its negative connotation, creates a conflict of interest for the researcher and renders misinformation a highly value-laden concept (Kharazian et al., 2024; Uscinski, 2023; Williams, 2024; Yee, 2023). Fourth, due to the context dependency in identifying misinformation (and limitations in data), there is no reliable method to quantify misinformation, false beliefs, and misinformation effects (Southwell et al., 2022). This limitation means we cannot answer straightforward questions such as "Are we seeing less or more misinformation or misinformation effects over time?" For example, many misinformation experts believe we live in a post-truth era (Altay, Berriche, Heuer, et al., 2023), but we do not have a method to verify it.

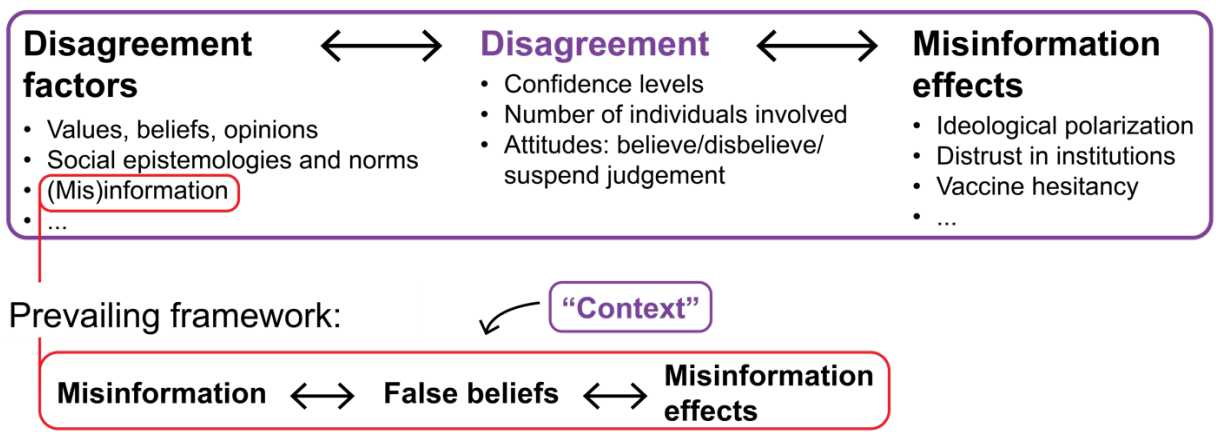

*Figure 1. Disagreement vs. misinformation frameworks. When studying misinformation effects like vaccine hesitancy and ideological polarization, the presumed causal pathway is that false and misleading information about factual matters consumed at the individual level negatively influences beliefs, leading to these effects. However, the formation of false beliefs and the associated effects are complex, involving a range of interdependent factors beyond just misinformation. The prevailing concept, which focuses on the spread of misinformation, overlooks these broader elements, despite recent research highlighting their significance. In contrast, disagreement is a more holistic concept that explicitly includes these normative and societal factors, offering better explanations for misinformation effects and guiding the development of more effective mitigation strategies.*

Given the limitations of the prevailing misinformation concept, developing a more effective framework for understanding how misinformation effects emerge, evolve, and can be mitigated is essential. Here, we propose framing misinformation effects as symptoms (and effects) of social disagreement, characterized by conflicting attitudes among individuals and communities. First, disagreement is a holistic concept that aligns better with misinformation effects. As a consequence, disagreement is necessary for these effects, while the prevalence of misinformation itself or false beliefs is not. Second, disagreement is shaped by disagreement factors, such as values, beliefs, and epistemologies, which are consistent with factors recent misinformation research has identified as significant contributors to misinformation effects (e.g., Altay, Berriche, & Acerbi, 2023; Aghajari et al., 2023), but they are inherently excluded by the prevailing concept. Third, disagreement acknowledges that individuals with differing values and epistemologies may reach different conclusions about the truth, requiring researchers to assess when disagreement causes harm. Like parents who may not focus on who is right or wrong when their children argue, a disagreement researcher can ignore the topic and still study the social dynamics and effects. Fourth, due to its conceptual alignment and because its identification does not require contextual analysis, disagreement offers a more practical way to measure (the risk of) misinformation effects, which can be fully automated, as we demonstrate. In this paper, we argue that, compared to misinformation, disagreement can offer greater explanatory power in analyzing misinformation effects, increased productiveness in developing new intervention strategies, and improved predictive precision in quantifying these effects (see "cognitive values" in Douglas, 2009), as we explain in the following three subsections. Additionally, in the Findings section, we demonstrate how this framework addresses three key research questions: 1) Why are Republicans more susceptible to misinformation than Democrats? 2) How can we counteract effects like vaccine hesitancy and ideological polarization? and 3) Do we live in a post-truth era? While the main focus of our paper is conceptual, our quantitative findings demonstrate that the proposed disagreement framework can also be applied to quantitative analysis. Due to space constraints, these quantitative contributions are detailed in the Appendix.

The purpose of this paper is neither to challenge misinformation research nor to discourage researchers from adopting a normative stance. Rather, since disagreement encompasses and incorporates misinformation as a factor, it can serve as a bridging concept (Chadwick & Stanyer, 2022) between in-depth, localized research focused on the spread of misinformation (Pasquetto et al., 2024) and research focused on its effects. Our suggestion to focus on disagreement should not be mistaken for epistemological relativism (i.e., the idea that knowledge is only valid relative to a specific social or cultural situation), which is problematic as it would prevent researchers from making objective claims (Douglas, 2009; Longino, 2002). While researchers often rely on weak objectivity when assessing whether a text negatively affects someone's beliefs, there is commonly strong scientific evidence to label the associated beliefs and behaviors as false—such as racist attitudes or vaccine refusal. Thus, not all positions within a given disagreement are equal. However, as we argue in this paper, even when scientific evidence clearly supports one side of a disagreement, the proposed framework remains valuable for understanding and mitigating its effects. At the end of this section, we provide recommendations for researchers on how to navigate these challenges.

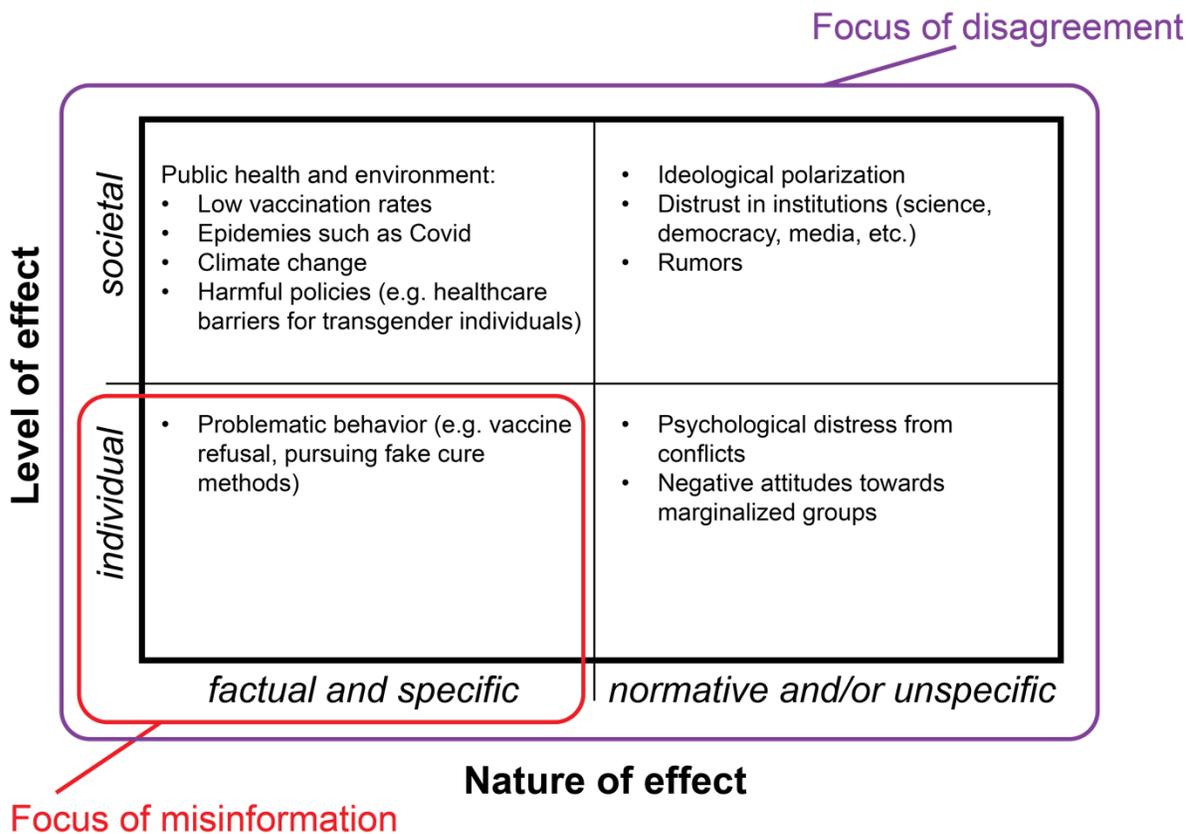

**Figure 2. Misinformation effects: social, political, and psychological issues associated with misinformation.** *These effects involve both factual and normative matters (nature of effect) and involve agencies at both individual and societal levels.*

*Disagreement provides better explanations for misinformation effects*

By elevating the conceptual focus from the content of information at the individual level (misinformation) to the overarching epistemic tensions between different communities (disagreement), the analytical framework aligns more effectively with the misinformation effects, see Figure 2. Because disagreement inherently recognizes the significance of normative and societal elements (see Figure 3) and of conflicting attitudes when studying the misinformation effects, it provides distinct explanations when compared to the concept of misinformation, see Table 1. For example, if one is interested in analyzing the influence of (fake) experts on vaccine hesitancy, the misinformation concept requires researchers to explicitly include such interdependent factors (Harris et al., 2024). In contrast, the holistic disagreement framework inherently incorporates these elements, compelling researchers to compare and evaluate the significance of individual factors for the misinformation effects under study. Taking factors such as values, identity, and epistemologies into consideration, it becomes less surprising why individuals may arrive at conclusions about given data and believe certain ideas that scientists deem misinformation (Douglas, 2009; Furman, 2023). For instance, misogynists may believe stories that target women, racists may believe stories that target immigrants, or Republicans may be more susceptible to misinformation than Democrats due to their stronger disagreement with science as an institution and its members (see

Conceptual Contribution 5). Understanding the strength, nature, and subject matter of disagreement can help analyze these problems and develop intervention strategies.

# Disagreement factors

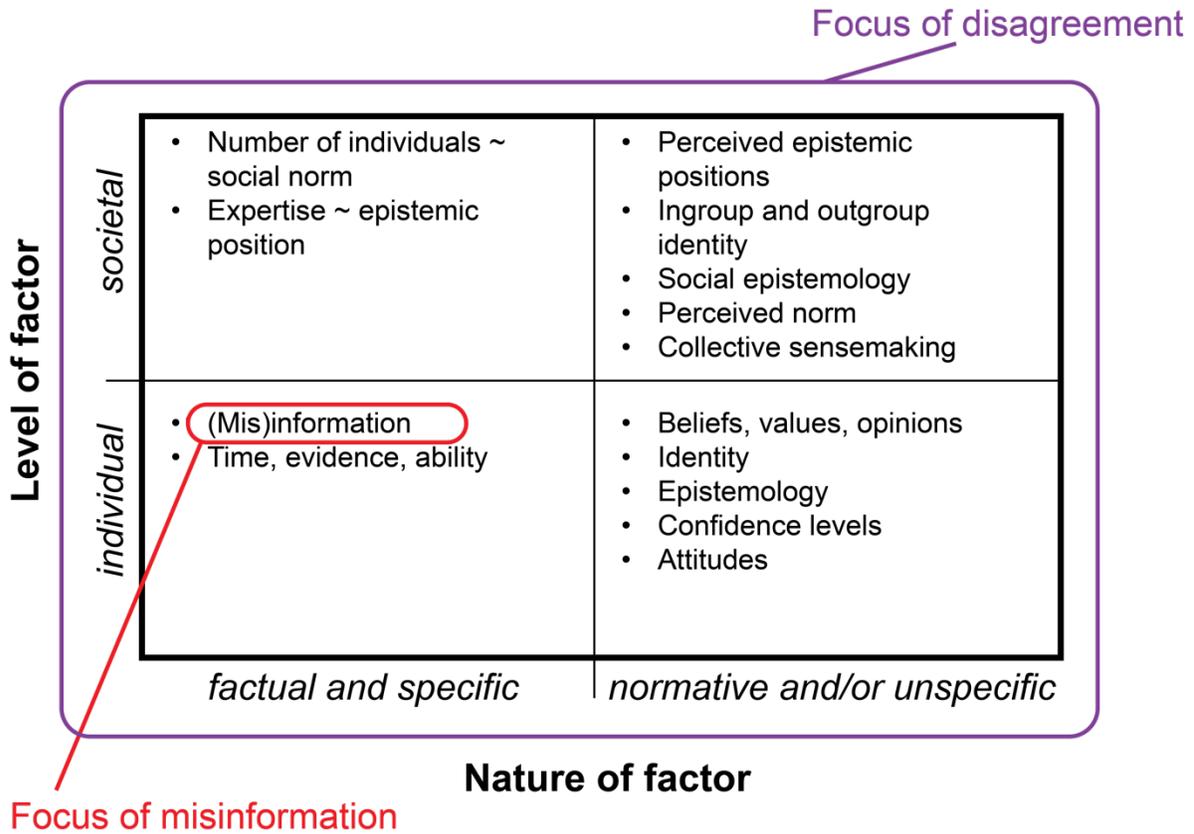

Figure 3. **Disagreement factors**. *Reflected in its factors, disagreement inherently incorporates normative and societal elements that prior work has identified as direct or indirect contributors to misinformation effects but overlooked in the prevailing misinformation concept.*

**Table 1.** *Comparison of misinformation and disagreement concepts along guiding questions.*

| | Misinformation | Disagreement |
|---|---|---|
| **Conceptualization** | | |
| **Scope** | Factual matters often at individual level, inherently topic-specific. | Broad spectrum: factual and normative matters, at individual and societal levels, topic-specific and topic-agnostic. |
| **Relationship with effects and associated harm** | Misinformation as a cause of misinformation effects. For example, assumption that misinformation about vaccines makes individuals reject vaccines. | Associated harm depends on the matter, nature, and strength of the disagreement. Disagreement strength as a risk indicator for misinformation effects. Disagreement factors as cause of disagreement. |
| **Connotation of concept** | Inherently negative which creates conflict of interest for researchers. | Neutral. |
| **Role of researcher** | Debater and judge, often required to make (normative) judgments regarding *misleadingness* or *truth*. | Can take an independent role studying the phenomenon in a topic-agnostic manner. |
| **Guiding questions and approaches** | | |
| **Analyzing effects** | <ul><li>What is considered misinformation?</li><li>What is the spread of misinformation?</li><li>Why do people share and believe misinformation?</li><li>What factors increase/decrease the spread of misinformation?</li></ul> (Underlying assumption that misinformation is inherently negative) | <ul><li>What communities are disagreeing about what matter (specific/unspecific)?</li><li>Why are they disagreeing?</li><li>What is the strength of disagreement? (numbers, confidence levels, epistemic positions)</li><li>Which disagreement factors are most relevant for the studied effects?</li><li>To what extent will disagreement about the topic raise the potential for individual, social, and institutional harm?</li></ul> |
| **Countering effects** | <ul><li>How to prevent the spread of misinformation?</li></ul> | <ul><li>What is the healthy extent of disagreement about the given matter?</li><li>What disagreement factors can be modified to achieve ideal disagreement?</li><li>How to disagree better?</li></ul> |
| **Measuring effects** | Counting instances of misinformation and surveys on false beliefs. | Disagreement in text data (e.g., sentiment analysis) or surveys, either directly or by analyzing attitudes (beliefs, values) for conflicts. |

*Note: Unlike misinformation, disagreement is not considered inherently negative and compels the researcher to evaluate when and where disagreement about a given topic results in associated harm.*

*Disagreement identifies interventions against misinformation effects*

Being one of many drivers, addressing misinformation alone will not eliminate the misinformation effects. Despite the significance of other factors, there are relatively fewer studies that focus on interventions at the societal or normative level, such as leveraging social norms and their perception to address collective action problems (Aghajari et al., 2023; Andı & Akesson, 2021; Nyborg et al., 2016) or increasing trust in institutions (Acerbi et al., 2022).

Because the disagreement framework asks distinct questions (see Table 1), it can help identify new interventions for addressing misinformation effects. Additionally, as a holistic framework, it facilitates the balancing of a combination of interdependent strategies, recognizing that individual interventions may have limited impact (Altay, Berriche, Heuer, et al., 2023; Bak-Coleman et al., 2022). To illustrate, we first reevaluate nine well-studied intervention strategies (Kozyreva et al., 2024) using the disagreement framework (Conceptual Contribution 6). We find that four of these strategies may be less effective because simply being aware of a disagreement does not necessarily change one's belief (Kelly, 2005). Second, we demonstrate how the questions from Table 1 can be used to develop and balance (new) intervention strategies (see Conceptual Contribution 7). For vaccine hesitancy, the proposed framework prompts questions such as "What are the reasons for individuals to disagree with governmental recommendations to get vaccinated?" and "How can individuals be motivated to get vaccinated while accepting their disagreement with governmental recommendations?" At times, calling out misinformation can help align beliefs or actions. However, considering disagreement factors, strategies like focusing on shared goals may be more effective (Friedman & Hendry, 2019). For example, instead of labeling vaccine misinformation, social media platforms could identify conflicting posts, which we argue is easier to automate than detecting whether a post is true or not and then append those conflicting posts with notes emphasizing shared health goals such as: "We observe the circulation of rumors about vaccines. We all want to save lives."

*Disagreement measures misinformation effects*

In January of 2024, the World Economic Forum (WEF) identified misinformation and its effects as one of the top risks faced in this year (World Economic Forum, 2024). How can we determine the long-term effectiveness of intervention strategies, assess whether emergent technologies like artificial intelligence (AI) or social media have increased the problem, or compare its progression between countries (Humprecht, 2019)?

Due to its subjectivity and context dependency, identifying misinformation requires human judgment and topic analysis (see Conceptual Contribution 2). As a consequence, measuring misinformation can only be done for known narratives and requires significant effort (Southwell et al., 2022). Even approaches leveraging AI rely on human-labeled datasets used for pretraining such models. These limitations contrast sharply with the considered concern and rapid dissemination. Moreover, when viewing misinformation as a general, unspecific problem—for example, the concern that we live in a post-truth era (Altay, Berriche, Heuer, et al., 2023)—there is currently no approach to track its change over time other than relying on the opinions of experts (Altay, Berriche, Heuer, et al., 2023).

Quantifying the (risk of) misinformation effects through disagreement strength instead of counting misinformation comes with two key advantages: First, its independence of context allows to automate measurements across topics and times; in particular, it enables capturing misinformation effects as a general problem, including effects stemming from unknown or emerging narratives (e.g., polarization). Second, if misinformation effects are more accurately captured by the social phenomenon of disagreement than by instances of misinformation, as we argue here, then disagreement serves as a better indicator.

To demonstrate the first advantage, we conducted a longitudinal disagreement measurement in text in letters to the editor of *The New York Times* (NYT) from 1950–2022 based on a lexicon-induced (negative) sentiment analysis approach (see Quantitative Finding 1). We found an increase in disagreement since 2006, which aligns with the prevailing assumption that we live in the post-truth era (Altay, Berriche, Heuer, et al., 2023). A large portion of the dataset used (1950–2007) originates from a prior study on conspiracy theories (Uscinski & Parent, 2014). While our disagreement measurement did not require any human labeling, the authors of the original study manually classified each letter to determine whether it contained misinformation (engagement with conspiracy theories).

Due to the lack of ground truth data for misinformation effects, we can only provide preliminary quantitative evidence that disagreement serves as a better indicator. In Appendix B, we include a comparison between disagreement and misinformation measurements for two longitudinal text datasets (letters to the editor of the NYT and vaccine-related tweets on Twitter). For both datasets, our analysis reveals a significant correlation, providing preliminary evidence that disagreement and misinformation are associated and may approximate the same effects. Furthermore, we found that vaccine-related disagreement decreased between January and March 2021 and increased between April and September 2021, confirming vaccine hesitancy analyses by Liu & Li (2021) and Chang et al. (2024), respectively. More data on misinformation effects is needed to validate whether disagreement serves as a better indicator of these effects than simply counting instances of misinformation or related concepts. Likely, the ideal measure depends on the research endeavor and might even be a combination of various approaches. Disagreement offers a framework for longitudinal, cross-national, and topic-agnostic analyses of misinformation effects, addressing a need identified by experts (Altay, Berriche, Heuer, et al., 2023).

In addition to social media and letters to newspaper editors, the disagreement measurement approach can be applied to other text data where individuals express their attitudes, enabling us to track misinformation effects in both specific and broader contexts. For example, it can be used with Wikipedia edit histories, commentary on government websites, reviews of health-related books, online forums, and comments on news sites, among others. Additionally, measuring unspecific disagreement can serve as a warning system for emerging misinformation effects. Instead of text analysis, disagreement can be analyzed through surveys, either by directly assessing perceived disagreement or indirectly through questions about values, beliefs, and interests (Voelkel et al., 2024; Akiyama et al., 2016).

*Recommendations for misinformation researchers and practitioners*

Identifying and studying misinformation is complex and context-sensitive. Some scholars argue that misinformation should not be studied at all (Williams, 2024), while others call for clear, context-free definitions (J. Uscinski et al., 2024). We propose a different approach: incorporating context by studying the effects within the overarching disagreement framework. Therefore, the purpose of this paper is neither to challenge misinformation research nor to discourage researchers from taking a clear stance within a disagreement. Compared to evaluating negatively connoted misinformation,[5] determining whether two communities are in disagreement requires less evidence and is less value-laden, allowing for analysis without normative judgment. This neutral perspective is particularly relevant for studying misinformation effects that do not fit neatly a simple true/false dichotomy, such as polarization and rumors, or when access to data and communities is insufficient to assess whether specific information is misleading—such as in longitudinal, cross-community, and cross-topic analyses and measurements. At this high level, studying disagreement is like studying the climate (e.g., earth temperature). Both are inherently neutral phenomena, but they can become problematic when they change or reach certain levels, making it crucial for researchers to understand how their intensity relates to potential harm. Even when researchers focus on the spread of misinformation itself—rather than its effects—the disagreement framework can still add value by capturing the overarching social dynamics and key factors that explain why individuals believe and share information that researchers evaluate as misleading. Additionally, disagreement serves as a bridging concept (Chadwick & Stanyer, 2022) between in-depth, localized research (Pasquetto et al.,2024) and broader, high-level analyses, as well as with related disciplines such as philosophy, computational linguists, psychology, and political science (e.g., Christensen, 2007; Lamers et al., 2021; Rosenthal & McKeown, 2015; Weinzierl et al., 2021; Landemore, 2017). Social crises drive social resistance, inevitably resulting in disagreement, but often also leading to rumors, false beliefs, and misinformation (Prooijen & Douglas, 2017). Disagreement can drive positive change (e.g., women's rights, Black Lives Matter) or negative outcomes (e.g., distrust in democracy). Depending on scientific evidence and researcher's normative stance, the goal may be either to resolve the disagreement independent of absolute positions or to bring the disagreeing community into alignment with the researcher's epistemology. When taking a stance against claims (e.g., "vaccines cause autism") or communities' actions (e.g., vaccine hesitancy), we recommend that researchers explicitly define the potential harmful consequences of these claims or actions. Clearly articulating these harms can help justify the need to bring the disagreeing community into alignment with the researcher's position.

---

[5] It is important to note that not all misinformation has a negative impact, despite its negative connotation. For example, falsely claiming a sports event like the Olympics occurred in a different location may not cause harm.

## Findings

*Conceptual Contribution 1: Conceptual misalignment between the narrow focus of misinformation and the broad spectrum of misinformation effects.*

The focus on misinformation and its underlying causal pathway (misinformation—individuals holding false beliefs—misinformation effects) does not account for the temporal and positional relativity of *truth* and fails to capture the normative and societal contributors to misinformation effects. For certain misinformation effects such as polarization or rumors, the current concept is not merely narrow but also misdirected. Most obviously, excluding the individuals holding *true* beliefs overlooks the impact that conflicting viewpoints (from experts and peers) have on our understanding of knowledge and belief. Adopting an attitude in favor of specific (mis)information not only positions oneself *for* a particular community but also *against* the opposing one (Pereira et al., 2021).

Figure 2 categorizes misinformation effects based on their matter (factual and normative) and level of occurrence of the underlying social dynamics (individual and societal). For example, ideological polarization is defined as "a situation under which opinions on an issue are opposed to some theoretical maximum" (Au et al., 2022, p. 1331). By definition, this corresponds to an effect involving normative matters (opinions) at the societal level, which does not necessarily require the presence of misinformation or false beliefs. While effects like problematic behavior due to false beliefs can occur at the individual level, effects like polarization, rumors, and environmental issues only manifest when a large number of people are involved. Moreover, many effects, such as rumors, polarization, distrust, and psychological effects (e.g., distress and discomfort) (Ilies et al., 2011; Rocha et al., 2023; Susmann & Wegener, 2023) do not primarily hinge on individuals holding false beliefs but rather on individuals or communities holding conflicting attitudes. These misinformation effects become problematic only when associated activities cause harm. For instance, polarization or distrust in institutions can be beneficial to a certain extent. The concept of misinformation, often seen as inherently negative, fails to acknowledge these nuances.

Given the broad range of social, political, and psychological phenomena representing misinformation effects, it becomes evident that interdependent factors beyond misinformation significantly influence these effects (Southwell et al., 2020). For example, a survey among misinformation experts highlights partisanship, identity, confirmation bias, motivated reasoning, and distrust in institutions as major contributors (Altay, Berriche, Heuer, et al., 2023). While sometimes acknowledged as context, they are often only peripherally considered as contributors to the spread of misinformation rather than directly to its effects. Our literature review suggests that these factors are as crucial as misinformation itself and should be directly incorporated into analyses. Figure 3 categorizes these factors by their nature and level of occurrence. They include opinions (Ancona et al., 2022; Facciolà et al., 2019; Koudenburg & Kashima, 2022; Niu et al., 2022), the epistemic position of individuals who hold true or false belief such as experts (Bongiorno, 2021), fake-experts (Harris et al., 2024; Lewandowsky et al., 2017; Schmid-Petri & Bürger, 2022), and influencers in general (Lofft, 2020), dispositional beliefs (Altay, Berriche, & Acerbi, 2023), the framing of information and experiences (Goffman, 1974; Starbird, 2023; Zade et al., 2024), ingroup and outgroup identity (Altay, Berriche, Heuer, et al., 2023; Pereira et al., 2021; Van Bavel et al., 2024), number of people involved in disagreement related to social norms and perceived norms (Aghajari et al., 2024; Andı & Akesson, 2021; Andrighetto & Vriens, 2022), shared and conflicting values and aesthetics (Aghajari et al., 2023), attitudes (Susmann & Wegener, 2023; Xu et al., 2023), social epistemologies (Bernecker et al., 2021; Furman, 2023; Pennycook & Rand, 2019; Uscinski et al., 2024; Wakeham, 2017), trust in institutions (Humprecht, 2023), disagreeing views (Mazepus et al., 2023; Uscinski, 2023) and so on.

*Conceptual Contribution 2: Identifying misinformation relies on human judgment.*

The conceptual misalignment has expanded the focus from scientifically false information to misleading information. However, whether and how information changes one's belief is not self-evident but context-dependent and somewhat subject to researchers' judgment. Consequently, there is no reliable, generalizable procedure to identify misinformation (Uscinski, 2023), limiting the approaches to automate quantification of misinformation and its effects (Southwell et al., 2022). Furthermore, in combination with the negative connotation of misinformation, it can create a conflict of interest for researchers (Uscinski, 2023) and epistemic injustice in misinformation detection (Neumann et al., 2022).

*Conceptual Contribution 3: Disagreement aligns with misinformation effects.*

In alignment with misinformation effects, disagreement is shaped by disagreement factors, involving both factual and normative matters, like opinions, values, and beliefs (Frances & Matheson, 2018; Furman, 2023) and can occur at both individual and societal levels. Unlike misinformation, it acknowledges the significance of conflicting attitudes in studying and mitigating misinformation effects (Mazepus et al., 2023), making it a necessary condition for these effects.

Disagreement occurs when at least two individuals adopt conflicting doxastic (belief-related)[6] attitudes toward the same content (Zeman, 2020). These attitudes—belief, disbelief, and suspension of judgment—can vary in confidence. The strength of disagreement depends on the disparity in attitudes, confidence levels, and the number of people involved, serving as a risk indicator for misinformation effects. Disagreement factors, which create and influence disagreements, correspond with contributors to misinformation effects (see Figure 3). Misinformation can be seen as a disagreement factor that corresponds to evidence (Frances & Matheson, 2018).

---

[6] Other types of conflicting attitudes can be converted to a doxastic disagreement (Zeman, 2020).

*Conceptual Contribution 4: Disagreement is necessary for misinformation effects, while misinformation itself is not.*

In the prevailing concept of misinformation, individuals holding false beliefs link misinformation to its effects. However, false beliefs cannot exist without true beliefs held by others, leading to a disagreement between those who believe the misinformation and those who do not—forming two disagreeing communities. Typically, researchers align with the true-belief community. Instead of dividing by false/true beliefs, communities can also be split between those benefiting from and those harmed by the misinformation. For example, the claim that vaccines cause autism aligns with the views of those who believe vaccines are unsafe. These divided communities are crucial: misinformation only matters if some believe it and seek to persuade others to believe it (i.e., disinformation), while new ideas accepted by everyone and in particular the scientific community are considered knowledge gain, not false belief. Hence, disagreement is necessary (but not sufficient) for false belief and the misinformation effects.[7]

*Conceptual Contribution 5: Republicans are more susceptible to misinformation due to their disagreement with science.*

Studies indicate that Republicans are more susceptible to misinformation than Democrats (Pennycook & Rand, 2019; Pereira et al., 2021). The disagreement framework may explain this difference. Differences in values, beliefs, and epistemic methods lead to varying conclusions about which information is relevant, which questions to ask, and how to interpret data (Douglas, 2009; Furman, 2023; Longino, 2002). Since Republicans tend to disagree more with the scientific community than Democrats (Evans & Hargittai, 2020; Lee, 2021), we can expect Republicans to differ more frequently from scientists in assessing information veracity. Conversely, if individuals with background assumptions more aligned with Republicans were responsible for distinguishing between true and misleading information, Democrats could be perceived as more susceptible to misinformation. However, we do not argue that misinformation or false beliefs are arbitrary or unworthy of study, as discussed in the Implications section.

*Conceptual Contribution 6: Disagreement framework reveals limitations of current intervention strategies.*

We reevaluate nine main intervention strategies from a recent high-profile paper (Kozyreva et al., 2024) viewed within the disagreement framework. Four of these strategies (row 3 in Table 2) may be less effective because they merely inform individuals that other individuals, communities, and institutions (e.g., science, media, government) disagree with their beliefs. However, simply being aware of a disagreement does not necessarily change an individual's belief (Kelly, 2005; Southwell et al., 2019); it largely depends on their epistemic relationship with the disagreeing community (Frances, 2014). If individuals do not perceive peers, fact-checkers, scientists, or the majority (social norm) as epistemically superior (better positioned for judgment), it is reasonable for them to remain steadfast in their beliefs. Thus, if individuals disagree with science, such as on the safety of vaccines, there are broadly three possible explanations: they are not aware of the scientific consensus on vaccine safety, they do not perceive scientists as epistemic superior, or they do not respond to disagreement in a reasonable way—unreasonable from a scientific perspective. Regardless of someone's relationship with the disagreeing community, intervention strategies like media literacy tips can help reduce disagreement by aligning individual epistemologies with scientific epistemology.

---

[7] False belief is necessary but not sufficient for misinformation.

**Table 2.** *Nine main misinformation interventions (Kozyreva et al., 2024) recast and reevaluated applying the disagreement framework.*

| Misinformation intervention type | Disagreement framework (disagreement factors in bold) |
|---|---|
| Accuracy prompts, friction, lateral reading and verification strategies. | Providing additional **time** and nudging individuals to collect more **evidence**. |
| Media literacy tips. | Alignment of **epistemology**, increase in **ability** to make a judgement. |
| Inoculation, warning and fact-checking labels, source-credibility labels, debunking and rebuttals, e.g., stating the truth, social norms. | These strategies largely make individuals aware of a disagreement, which does not necessarily change one's beliefs. |
| Debunking and rebuttals, e.g., offering explanation to the data. | Providing additional **evidence**. |

*Note: Similar interventions are grouped together, while the "Debunking and rebuttals" intervention is split into two rows, as its evaluation with the disagreement framework depends on the specific execution.*

*Conceptual Contribution 7: Disagreement offers new intervention strategies exemplified for vaccine hesitancy and ideological polarization.*

Mitigating the spread of misinformation about vaccines and other factual matters *can* help decrease effects such as vaccine hesitancy or ideological polarization. However, there are many other reasons beyond false beliefs why individuals might disagree with governmental recommendations to get vaccinated or why societies might develop strongly opposed opinions (Au et al., 2022). In Appendix A, we demonstrate how the proposed disagreement framework, along with the related guiding questions and measurement approaches (see Table 1), can help identify and balance intervention tactics.

*Quantitative Finding 1: Disagreement increased since 2006.*

Disagreement offers a way to measure the risk of misinformation effects. Based on the widely accepted idea that we live in the age of misinformation (Altay, Berriche, Heuer, et al., 2023), we expect general, unspecific disagreement to be higher in the current decade compared to previous times. Figure 4 shows the time series of yearly disagreement scores from 1950–2022 in letters to the editor of the NYT. Disagreement has increased since 2006, providing preliminary evidence for the idea that effects of misinformation have increased in the past years. We used an automatic and scalable approach based on lexicon-induced (negative) sentiment analysis to measure disagreement in the text of letters to the editor of the NYT. Disagreement was measured by averaging the negative sentiment scores of about 1,000 letters per year, normalized to a scale between 0 and 1 across all years. For the years 1950–2007, we used the letters dataset from Uscinski & Parent (2014) to compare with their manually derived yearly misinformation score. To extend the coverage to 2022, we expanded the dataset with letters accessed through ProQuest.[8] The approach is detailed in Appendix B along with a longitudinal disagreement measurement of vaccine-related tweets.

---

[8] https://www.proquest.com/

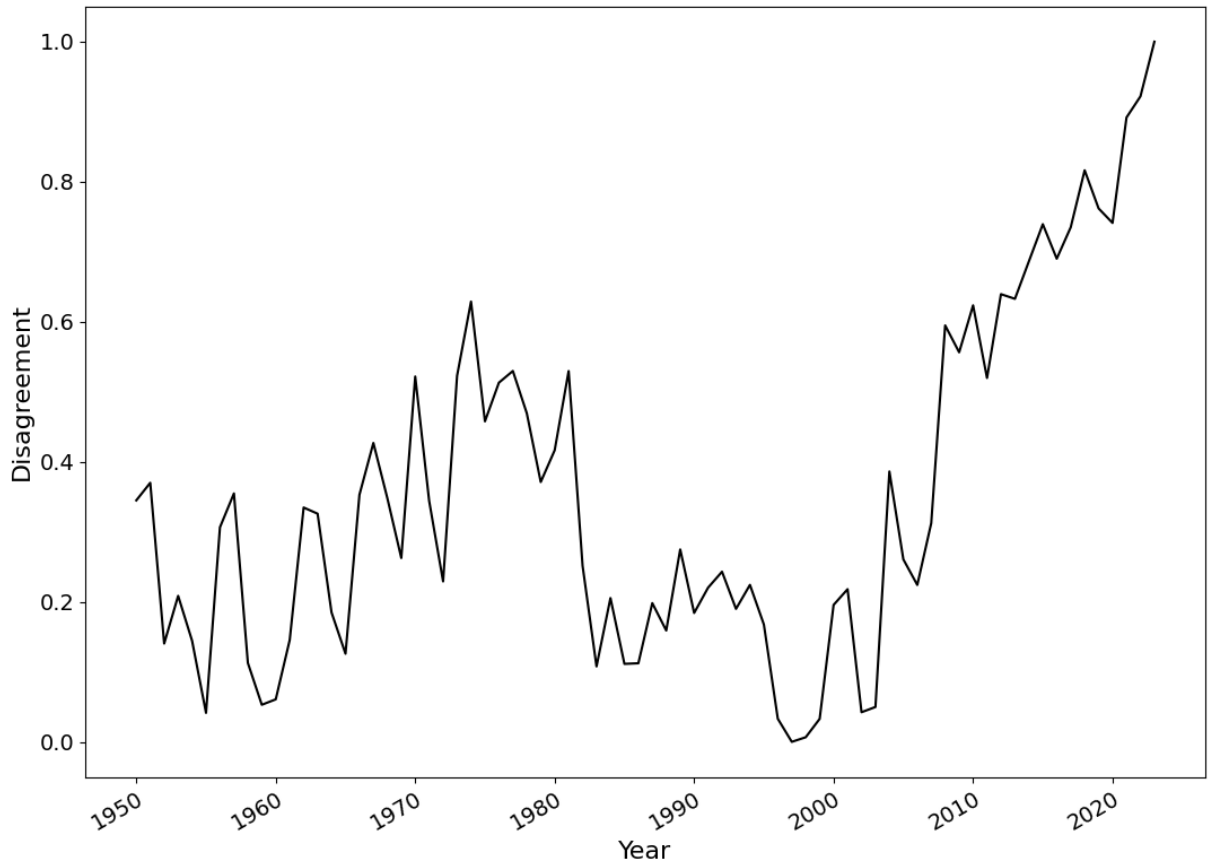

***Figure 4. Disagreement increased since 2006.*** *Normalized disagreement score in the letters to the editor of the NYT from 1950 to 2022 for about 1,000 letters per year.*

## Methods

As misinformation researchers ourselves, we grapple with the issues of the prevailing analytical framework, as raised by scholars both within and outside the field (Pasquetto et al., 2024). Aiming to address the critique and to move the field forward, we chose the following approach. First, in reviewing the literature on misinformation, we aimed to precisely identify the gaps between definitions and theories of misinformation (causes) and the problems under study (effects). Second, we reviewed related literature on belief formation and epistemology to synthesize an analytical framework that aligns better with misinformation effects. Finally, by comparing the two frameworks on three typical research questions (Conceptual Contributions 5 and 7, and Quantitative Finding 1), we aimed to demonstrate the effectiveness of the identified disagreement framework.

*Limitations and recommendations for future research*

Our argument for using disagreement over the misinformation concept is based on a fundamental comparison of common notions of both concepts. We define disagreement in a basic, intuitive way (Zeman, 2020), where at least two individuals hold different attitudes or confidence levels toward a specific proposition. We did not explore other notions of disagreement, leaving the ideal definition for studying misinformation effects to future research.

We use a conceptualization of misinformation that is narrow and only includes the kind where individuals believe it. Misinformation that doesn't include potential changes in belief is unlikely to spread widely in society or cause something like vaccine hesitancy. Once individuals believe the misinformation or actively seek to persuade others, it then can be captured by our proposed disagreement framework.

Building on our framework, we also introduce disagreement measurement as a risk indicator for misinformation effects. The outcome and interpretation depend heavily on the method and the social ecosystem analyzed. Our approach measures expressed disagreement in aggregated text data, which majority attitudes and neglects minority perspectives in society. Using negative sentiment analysis provides a rough estimate suitable for longitudinal comparison. We leave the development of more accurate measures and the exploration of alternative sources for analyzing disagreement to future research. Therefore, we view our empirical results as a demonstration rather than a benchmark for disagreement measurement.

**Acknowledgements**

We would like to thank Yiwei Xu and Anna Beers for their helpful feedback on early drafts of this paper.

**Funding**

This work is supported by the University of Washington's Center for an Informed Public and the John S. and James L. Knight Foundation (G-2019-58788), the IMLS (Grant # LG-255047-OLS-23), and the National Science Foundation (Grant # 2120496).

**Competing interests**

The authors declare no competing interests.

**Ethics**

The letters to the editor of the NYT is not owned by the authors (see Data availability). These letters have been published in the NYT and were partially used in prior work (J.E. Uscinski & Parent, 2014). The disagreement measurement was conducted at an aggregated level, and no personal data of the authors of these letters were used or inferred. The Twitter (now X) data is owned by the University of Washington's Center for an Informed Public (Starbird et al., 2023). This data was determined by the Human Subject Division at the University of Washington not to involve human subjects, as defined by federal and state regulations, and therefore did not require review or approval by the institutional review board (IRB). In accordance with Twitter's Terms of Service at the time of collection and to protect the privacy of Twitter users (Fiesler & Proferes, 2018), the only data we release are tweet IDs—this does not include tweet text, user IDs, usernames, media, or other fields. Analogous to the letters to the editor data, the disagreement and misinformation measurements were conducted at an aggregated level.

**Copyright**

**Data availability**

The letters to the editor of the NYT from "American Conspiracy Theories" (Uscinski & Parent, 2014) can be accessed by reaching out to its authors. All other materials needed to replicate this study are available via the Harvard Dataverse: https://doi.org/10.7910/DVN/UBLQPA. Additionally, the code is available on GitHub: https://github.com/hodeld/disagreement-misinfo.

# Appendix A: Intervening in vaccine hesitancy and ideological polarization

*Conceptual Contribution A1: Intervening in vaccine hesitancy.*

In case of vaccine hesitancy, there is strong scientific consensus on the safety and efficacy of vaccines. Misinformation conceptualizes vaccine hesitancy as a result of misinformation about vaccines. In contrast, our proposed framework reframes it as a disagreement about whether one should get vaccinated, between the majority who follow governmental recommendations based on scientific findings and individuals who disagree, for a range of possible reasons, see Table A1. Consequently, the main focus should be on minimizing disagreements on this matter. Viewing vaccine hesitancy as an issue of disagreement prompts a variety of analysis questions that allow to test new intervention tactics.

Table A1 lists potential questions for the analysis. Examples include: "What are the reasons for individuals to disagree with governmental recommendations to get vaccinated?"; "How can individuals be motivated to get vaccinated while accepting their disagreement with governmental recommendations?"; and "What are the common values and shared goals among both communities?" Addressing and correcting misinformation on the individual level is just one of several potential strategies. If individuals hesitate to get vaccinated due to distrust in science and government, publicly calling out misinformation is unlikely to decrease disagreement. Conversely, if hesitation is due to misinterpretation of data (epistemology), media literacy tips may be the right tool. If the primary issue is access to scientific information about vaccines, information campaigns could be the most effective approach. In cases of false dissensus, bolstering the credibility of scientists by showcasing consensus within the scientific community and informing about social norms may be beneficial.

Regardless of individuals' reasons for refraining from vaccination, fostering agreement between different groups by highlighting shared values and objectives, such as saving lives, can be a useful strategy. For example, social media platforms could append conflicting posts (identified using the disagreement measurement approach; see Figure 6) with notes emphasizing shared health goals and values among users (Friedman & Hendry, 2019). An example message could be: "We observe the circulation of rumors about vaccines. We all want to save lives." There is no definitive answer to the optimal strategy. The key point is that a variety of strategies must be carefully balanced, with a disagreement measurement as shown the next section could assist in evaluating newly identified interventions against vaccine hesitancy.

*Conceptual Contribution A2: Intervening in ideological polarization.*

Analogous to vaccine hesitancy, the misinformation framework conceptualizes polarization as a cause of misinformation, hence misleading information about factual matters. However, this concept is somewhat misguided, since polarization is defined as strongly opposed opinions (Au et al., 2022). In contrast, disagreement views ideological polarization as an extreme form of disagreement about any type of matter, as shown in Table A1.

The main goal of decreasing polarization is to ease tensions between opposing groups. Topic analysis, both online and offline, can help identify the matters of disagreement and the involved communities. One approach to counter online polarization could involve modifying the structure of platforms to decrease disagreement (Musco et al., 2018). Alternatively, posting notes stating, "We observe conflicting information on this topic, please be respectful to each other," may slow down the propagation of conflicting information and decrease polarization. The evaluation of such disagreement-based interventions is left for future research.

*Table A1*. Disagreement framework applied to ideological polarization and vaccine hesitancy.

| | Applying Disagreement to Ideological Polarization | Applying Disagreement to Vaccine Hesitancy |
|---|---|---|
| **Conceptualization** | Polarization is an extreme form of disagreement. | Vaccine hesitancy corresponds to the disagreement about whether one should get vaccinated, while the government, scientific community, and the majority endorse vaccination. |
| **Analysis** | • At what strength does disagreement become harmful?<br>• Between which communities do we see extreme disagreement?<br>• About what matter do they disagree the most?<br>• What disagreement factors are the main drivers of extreme disagreement? | • Who are the communities that disagree with the majority backed by science and government?<br>• Why do they disagree? (e.g., distrust in mainstream institutions, agreement with alternative medicine, fear, identity, misinformation)<br>• What is the strength of disagreement? (numbers, confidence levels, experts, fake experts)<br>• Who are the communities that agree with scientific recommendations and why do they agree? (trust, obedience, etc.)<br>• Is it acceptable to disagree with recommendations from science and government (from scientific, moral, and political perspectives)? |
| **Intervention** | • How can the main disagreement factors be controlled?<br>• What policies and social media designs can promote agreement?<br>• How to disagree better? | • How can individuals be encouraged to get vaccinated despite disagreement with government and science?<br>• How to increase trust in government and science?<br>• What are the agreements among different medical approaches?<br>• What are the shared goals and values among disagreeing communities? (e.g., preventing deaths)<br>• What other factors drive disagreement and how can they be controlled? (e.g., preventing the spread of misinformation) |

# Appendix B: Disagreement measurement

To illustrate how disagreement measurement can estimate the risk of misinformation effects, we conducted two longitudinal analyses: one on disagreement in letters to the editor of the NYT from 1950–2022 (Uscinski & Parent, 2014), and another on disagreement in vaccine-related posts on Twitter from 2020–2023. Computer linguistics offers a range of approaches to measure disagreement in texts (Augenstein et al., 2016; Bahuleyan & Vechtomova, 2017; Lamers et al., 2021; Misra & Walker, 2013; Rosenthal & McKeown, 2015). To demonstrate the advantages of measuring disagreement compared to misinformation quantification, we used an automatic and scalable approach based on lexicon-induced negative sentiment analysis.

To determine the negative sentiment of a given text, we calculated the average negative scores of words using SentiWordNet (Baccianella et al., 2010). The rationale for using negative sentiment as a proxy for disagreement and the method, including its validation against human-labeled sentiment scores, are detailed in the next two sections. The subsequent two sections detail how we measured disagreement in letters to the editor of the NYT and in vaccine-related Twitter posts. Finally, in Quantitative Finding 2, we validate the disagreement measurement approach as a method to estimate misinformation effects.

*Negative sentiment to automate disagreement measurement*

To automate misinformation detection and quantification, current approaches rely on machine learning systems trained on manually labeled texts containing misinformation (Southwell et al., 2022). Similar approaches exist for detecting and counting disagreement (Augenstein et al., 2016). However, such approaches are somewhat context-specific (Reuver et al., 2021) and cannot be directly applied to other languages, platforms, and text types. Therefore, we use a simple, lexicon-based approach measuring negative sentiment (Catelli et al., 2023; Wang & Cardie, 2016) as a proxy for disagreement. Although negative sentiment is not the most accurate approach for quantifying disagreement, it is most suitable for our purpose. Here is why: First, because we are not interested in detecting disagreement but in tracking changes in a given population over time, it is sufficient when our method converges to stable values in aggregated texts (Zollo et al., 2015). If the number of texts is large enough, even minimal precision is sufficient.

Second and related to the first point, although negative sentiment does only capture expressed negative attitudes (semantics of perspectival expressions) and thus, neglects many other forms of disagreement (e.g., between an individual who holds false belief and a person who suspends judgement), it especially captures disagreement about a matter of high stakes for both interlocutors, which generates conflict (Zeman, 2020). We assume that such disagreements are also more significant for problems studied in misinformation. When a matter is at stake, members of both opposing groups make claims that lead to negative attitudes and expressions from the other group. Let us provide an example in the context of vaccines. People holding a false belief adopt a negative attitude towards propositions such as "Vaccines are dangerous!" while individuals holding true belief disagree with claims such as "No, that's a lie!" or similar.

Third, although there are many other sociocultural factors than disagreement that affect negative sentiment, it is a suitable proxy for our purpose because many of them relate to the formation of disagreement and misinformation effects. Examples include polarization and the generation of so-called echo chambers (Humprecht et al., 2020), economic, cultural, and political crisis (Prooijen & Douglas, 2017), public concern (Ji et al., 2015), well-being (Cao et al., 2018), etc.

And last, our lexicon-based approach measuring negative sentiment is unlike deep learning approaches explainable (Hardalov et al., 2022) and generalizable as such sentiment lexica already exist for various languages (Shalunts & Backfried, 2015; Ucan et al., 2016) or could be created with relatively little effort. We leave the development of better approaches to measuring disagreement to future research.

*Negative sentiment method*

To estimate disagreement, we use negative sentiment scores based on SentiWordNet. SentiWordNet is a lexicon of more than 200,000 words along their positive, negative, and "objective" scores, explicitly created for sentiment analyses. SentiWordNet assigns each word in its corpus three values corresponding to negative, positive, and objective sentiment. These values range between 0 and 1 and collectively sum to 1. To determine the negative sentiment score of a given text, we use SentiWordNet to calculate the average score. This is done by summing all the negative scores of the words in the text and then dividing by the total number of words. The equation below shows the formula for calculating disagreement score $d$ for a given text with $N$ words contained in SentiWordNet. $Neg$ corresponds to the negative sentiment score of the text and $neg_i$ the negative sentiment score for a given word obtained from SentiWordNet.

$$d = Neg = \frac{\sum_{i=0}^{N} neg_i}{N}$$

We validate our method for both negative sentiment classification and disagreement classification. For negative sentiment classification, we use the IMDB review dataset (50,000 movie reviews; Maas et al., 2011) and achieve an F1 score and accuracy of 0.65. For disagreement classification, we use the UKP corpus, which contains more than 25,000 sentences along with their corresponding categories ("no argument," "support argument," "oppose argument"), and achieve an F1-score of 0.58 and an accuracy of 0.53. In both experiments, we convert the scores obtained through our method to binary labels by splitting the scores at the median. Correspondingly, we convert UKP categories to binary labels ("oppose argument" versus "no argument" or "support argument"). This approach allows us to compare the results with those of a random classifier, which would achieve an F1 score of 0.5.

*Disagreement in the letters to the editor of the NYT*

For disagreement in an unspecific context we use the data from (Uscinski & Parent, 2014) on letters to the editor of the NYT. As Uscinski & Parent (2014) note, letters to the editor of the NYT "are a solid source of national attitudes," (p. 58) making them ideal for measuring unspecified social disagreement. Furthermore, the selected dataset allows us to compare our disagreement measurement to misinformation scores derived from the same text data. While it would be reasonable to object to using published letters as a representation of social disagreement due to selection bias, the authors of *American Conspiracy Theories* (Uscinski & Parent, 2014) effectively address these concerns. For example, they validate the dataset by comparing letters to the editor from a newspaper with a very different political stance (*Chicago Tribune*) and provide statistical evidence that factors such as changing ownership, editorship, and circulation did not affect the data.

The dataset comprises about 1,000 letters each year. The authors generously provided the scans of the letters analyzed, with one PDF file for each year. Due to the low quality of the PDF scans prior to 1950, we only used data from the subsequent years (1950–2007). We expand the dataset with letters accessible through ProQuest[9] to finally cover the years 1950–2022. The final dataset comprises a total of 91,373 letters. To determine the yearly disagreement score, we average over a complete year because for the years prior to 2008 only yearly data are available.

*Disagreement in Twitter posts related to vaccines*

For the vaccine-related disagreement measurement, we used posts from Twitter (renamed X in 2023). Each month, between 2020 and 2023, we sampled 10,000 posts containing at least one of the keywords listed below. The final dataset comprised a total of 360,000 posts. To determine the monthly disagreement scores, we calculated the score for each post using the method introduced above and then averaged the scores over the 10,000 sampled posts per month.

Vaccine keywords:
%vacc%, %vax%, antivax, antivaxxer, c0v1d, c0vid, coronavaccine, coronavirusvaccine, covid passport, covid-19 passport, covid19vaccine, covidvaccine, exvaxxer, fauciouchie, gardasil, gardisil, hearthiswell, mandate, mandates, medical racism, new apartheid, novaccine, va((ine, va**ine, vaccinate, vaccinated, vaccination, vaccine, vaccine injury, vaccineinjury, vaccines, vaccinescauseautism, vaccinesharm, vax, vaxx, vaxxed, vaxxedII

*Quantitative Finding 2: Significant correlation between disagreement and misinformation.*

We introduced disagreement measurement as an approach to estimate the risk of misinformation effects. Validating such an approach is a non-trivial task due to the lack of ground-truth data, i.e., continuous, long-term analysis on misinformation effects. Therefore, we approximate them through misinformation.

For misinformation in an unspecific context, we used the data from (Uscinski & Parent (2014). For each year, the authors manually determined the proportion of letters engaging with conspiracy theories (Uscinski & Parent, 2014, p. 110) which serves as yearly misinformation score.

The second dataset consists of Twitter posts containing vaccine-related misinformation on Twitter from 2020 to 2022. To measure disagreement, we use the approach based on negative sentiment as introduced above. To classify the posts in misinformation/non-misinformation we use BERT[10] (Devlin et al., 2018) fine-tuned on a vaccine misinformation dataset (Hayawi et al., 2022) with the training parameters shown in Table B1. The proportion of 10,000 monthly sampled posts classified as misinformation serves as monthly misinformation score.

**Table B1.** *Training parameters for fine-tuning on vaccine misinformation dataset (Hayawi et al., 2022).*

| Parameter | Value |
| --- | --- |
| output_dir | model_dir |
| learning_rate | 2e-5 |
| per_device_train_batch_size | 16 |
| per_device_eval_batch_size | 16 |
| num_train_epochs | 3 |
| weight_decay | 0.01 |
| evaluation_strategy | "steps" |

---

[9] https://www.proquest.com/
[10] Uncased: https://huggingface.co/bert-base-uncased

| Parameter | Value |
|---|---|
| eval_steps | 500 |
| load_best_model_at_end | True |
| save_total_limit | 2 |

In both cases, we found a significant correlation of disagreement with misinformation, as indicated by the Prais-Winsten method. We used this method because it accounts for autocorrelation in time series data, and it was also employed by Uscinski & Parent (2014) to estimate correlations between time series in the same letters to the editor of the NYT. Figure B1 shows the time series of the engagement in conspiracy theories and aggregated disagreement scores of letters to the editor of the NYT. We found a significant Prais-Winsten correlation between the measures (p < .05). Figure B2 shows vaccine-related misinformation and disagreement in Twitter posts for a period of three years. The Prais-Winsten method reveals a significant correlation (p < .05). These results provide a preliminary validation of our approach measuring (the risk of) misinformation effects with disagreement.

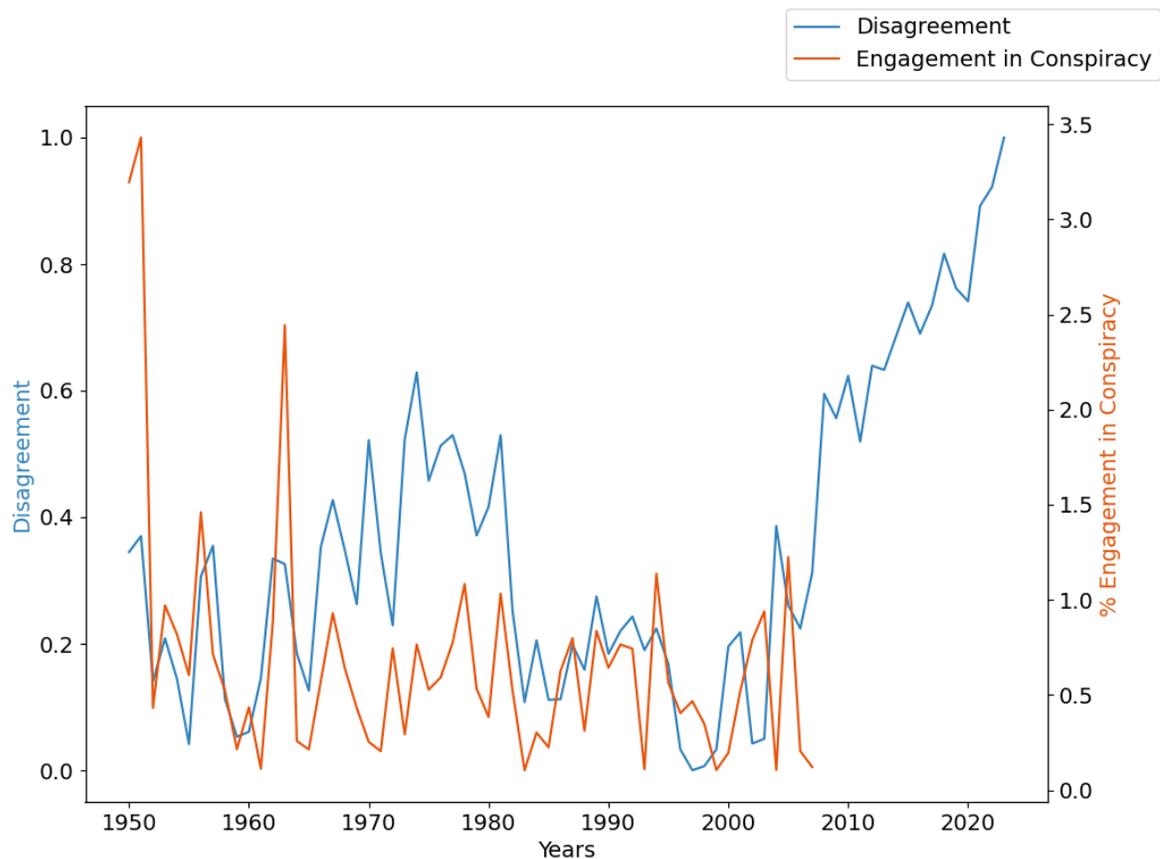

**Figure B1. Time series of yearly disagreement scores (blue) and engagement in conspiracy theories (orange) in the letters to the editor of the NYT.** *The two-time series show a significant correlation (p < .05) using the Prais-Winsten method. The disagreement reaches a global maximum in 2022. The engagement in conspiracy theory (1950–2007) corresponds to the proportion of letters containing such engagement; we used the data published by Uscinski & Parent (2014). We used the exact same letters for the disagreement analysis between 1950 and 2007. For the subsequent years (2008–2022), we used letters from the historical newspaper database ProQuest. The scores reflect an average of about 1,000 letters a year.*

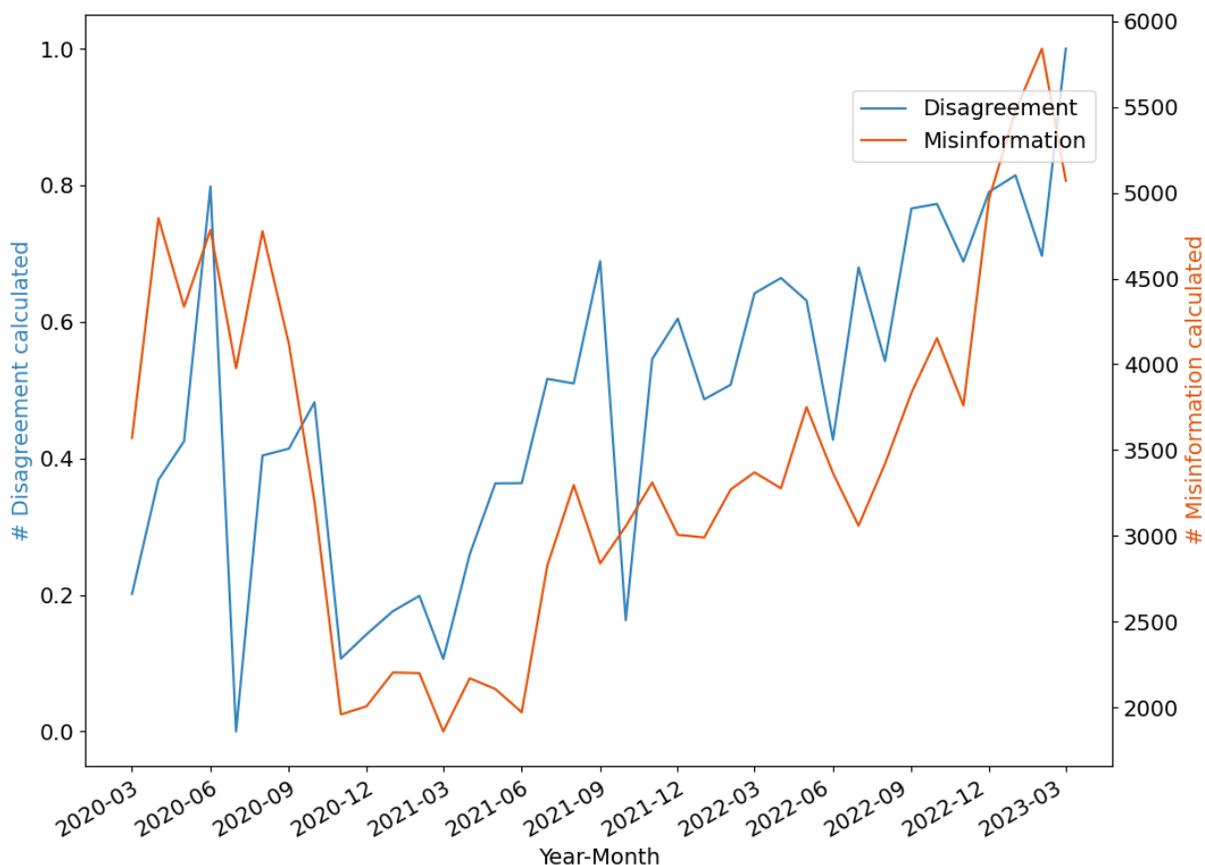

*Figure B2. Time series of the monthly aggregated disagreement scores (blue) in vaccine-related posts on Twitter and proportion of posts containing misinformation (orange) from 2020 to 2023 (36 months). The two series show a significant positive correlation (p < 0.05) using the Prais-Winsten method. The scores reflect an average of 10,000 posts a month.*

In our approach, we operationalized disagreement through negative sentiment as introduced above. Given that vaccine-related misinformation typically consists of negative sentiments against vaccines, this correlation is not surprising. Therefore, we also compared our analysis with survey data related to false beliefs about vaccines. Uscinski et al. (2022) analyzed various claims and found in general a drop in vaccine-related false beliefs between June 2020 and May 2021. Liu & Li (2021) and Chang et al. (2024) calculated vaccine hesitancy in the U.S. context and found a decrease between January and March 2021 and an increase between April and September 2021. In all three cases, our measured disagreement (obtained through negative sentiment) aligns with the rates of false beliefs and vaccine hesitancy over time. Unfortunately, there is limited temporal data on vaccine hesitancy or false beliefs available for comparison with our disagreement measure. This limited data availability highlights the challenge in obtaining such information and illustrates why measuring disagreement offers a valuable alternative for estimating the risk of misinformation effects. As demonstrated by the analysis of the letters to the editor, we also obtain a significant correlation between disagreement and engagement in conspiracy theories. Unlike vaccine misinformation, the conspiracy dataset covers a wide spectrum of false beliefs over a time span of more than half a century.